\documentclass[aps,pra,english,twocolumn,showpacs,preprintnumbers,amsmath,amssymb,floatfix,superscriptaddress,longbibliography]{revtex4-2}

\usepackage[utf8]{inputenc}
\usepackage[T1, T2A]{fontenc}
\usepackage{graphicx}
\usepackage{amssymb}
\usepackage{babel}
 \usepackage[font=small,skip=0.1in]{caption}
\makeatletter

\usepackage[version=3]{mhchem}

\usepackage{color}

\providecommand{\keywords}[1]
{
  \small	
  \textbf{\textit{Keywords---}} #1
}

\makeatother
\usepackage{babel}
\makeatother

\begin{document}

\title{``Observations on the possible electromagnetic nature of nucleon interactions and pions'' --- historical manuscript from 1969 by B. W. Ninham and C. Pask}

\author{Mathias  Bostr{\"o}m }
  \email{mathias.bostrom@ensemble3.eu}
 \affiliation{Centre of Excellence ENSEMBLE3 Sp. z o. o., Wolczynska Str. 133, 01-919, Warsaw, Poland}
 \affiliation{Chemical and Biological Systems Simulation Lab, Centre of New Technologies, University of Warsaw, Banacha 2C, 02-097 Warsaw, Poland}

\author{Drew F. Parsons}
\email{drew.parsons@unica.it}
\affiliation{Department of Chemical and Geological Sciences and CSGI University of Cagliari, Cittadella Universitaria, 09042 Monserrato, CA, Italy}

\date{\today}%

\begin{abstract}
 This manuscript presents an historical perspective prepared by Barry Ninham and  Colin Pask in 1969 on the connection between quantum electrodynamics theory and nucleon interactions. 
 A new theory of strong interactions based on electromagnetic considerations is proposed. Energy and force range magnitudes are correctly given. A new theory of the pion emerges and the pion mass and lifetime are calculated. No strong interaction coupling constant is required. 
\end{abstract}


\maketitle

 
\subsection{Historical preface by Bostr\"om and Parsons}

The general understanding \cite{PhysRevLett.99.022001} of nuclear and particle physics indicates, relying for instance on first-principle-lattice Quantum Chromodynamics simulations (QCD), that the main features of nuclear forces arising from the quark and gluon degrees of freedom can be described by the well-known QCD theory. The long-range behaviour of the nuclear force appears to be consistent with the pion-exchange potential \cite{PhysRevLett.99.022001}. Considering that these lattice QCD simulations usually neglect quantum electrodynamical interactions, it is usually assumed that electrons, positrons, and photons can be ignored when describing the main features of nuclear forces at the energy scale of pion exchange.
However, an original manuscript, written around 1969 by Prof. Emeritus Barry W. Ninham and Dr. Colin Pask (with minor edits during the following years), showed how mesons of a mass 220 times the electron mass could emerge from Casimir's original derivation of the forces between perfectly reflecting plates. They also derived the lifetime of the uncharged mesons. That was prompted by a conversation with Freeman Dyson who told Ninham that Feynman had believed there ought to be a connection between electromagnetic theory and nuclear interactions. This Feynman never found.  Later, Ninham and co-workers explored these links\,\cite{PhysRevA.57.1870,PhysRevA.67.030701,EPJDNinham2014,Ninham_Brevik_Bostrom_2022}. 
The force described by the Casimir effect\,\cite{Casi} is related to the ideas developed by Ninham and Pask. In general van der Waals, Casimir, and Lifshitz interactions can be linked to boundary conditions of the electromagnetic field at the interface between two uncharged or charged surfaces\,\cite{Casi,Dzya,ParNin1969,NinhamParsegianWeiss1970,Bost2000,Lamoreaux2012,PhysRevLett.120.131601}. These forces, in particular, have been measured\,\cite{TabNature,Tab,Lamo1997,Munday2009,SomersGarrettPalmMunday_CasimirTorque2018}, and recently for example predicted to impact the growth of ice at water-solid interfaces on planets, moons and exoplanets\,\cite{JohannesWater2019,BostromEstesoFiedlerBrevikBuhmannPerssonCarreteroParsonsCorkery2021,LiCorkeryCarreteroBerlandEstesoFiedlerMiltonBrevikBostrom2023}. Remarkably, they also act between protons and neutrons on the nuclear scale\,\cite{PanjaAnnPhys481_2025}. The interpretations presented in the work by Ninham and Pask may perhaps be considered controversial by some, as they indicate that certain aspects of the well-established quantum field theories in nuclear and particle physics can be described using a semi-classical electrodynamical theory. Ninham and Mahanty went on to develop a semiclassical theory of molecular dispersion forces \cite{Mahanty3,MahantyNinham_DispersionForces_book_1976}, capable of describing the Lamb shift \cite{Mahanty1974}, and at the heart of which lay the molecular polarisability. The Ninham-Pask theory of pion interactions presented here could be extended to the case of uncharged neutrons  by replacing the Klein-Gordon equation of eq.\eqref{eqn:KleinGordon} with Mahanty and Ninham's semiclassical electromagnetic vector equation, and introducing the neutron polarisability \cite{Thaler1959,Baldin1960,BaymBeck2016} arising from charged quark interactions \cite{LeeMilsteinSchumacher2002},  related to the fine structure constant \cite{MiorelliBaccaEtAl2016}.

We are finally able to present the original work by Ninham and Pask in this historical contribution. The following foreward originates from an interview with Ninham by Bostr\"om and Parsons (only slightly edited). The authors of the preface and foreword take no credit for the science in the main section of this manuscript, which we consider to be of historical interest. 
We present this hidden treasure ahead of Barry Ninham's 90th birthday (9 April 2026). The historical manuscript starts from Section\,\ref{intro}, ``Introduction''.

\subsection{Foreword from interview with Barry Ninham}

Barry Ninham wrote this previously unpublished paper after returning to Australia from USA in 1969. He has been then working with Adrian Parsegian at NIH  working on Lifshitz theory, which few at the time could properly understand. At that time Ninham invited Colin Pask to be a coauthor. At this time Colin Pask was  a young researcher working on nuclear physics -- which was what the manuscript was about.  After 55 years, Barry Ninham can  not fully remember the background as to why this manuscript remained unpublished. Most likely Pask considered it too controversial, and pulled out. This is an all too common response, people are often afraid of anything out of the ordinary or new. At that time it was thought that there was no way electromagnetic theory could have anything to do with nuclear forces. Thus Ninham got discouraged and dropped it. 
Colin Pask had a long career and later wrote a magnificent biography of Newton who went to the same school in Lincolnshire as Pask. It is called \textit{Magnificent Principia}\,\cite{pask2013magnificent}. Notably, Colin Pask was the first appointment Barry Ninham made when he took up his chair at Australian National University in 1971.

The concept of a quantum electrodynamic theory of nuclear interactions was later developed by Ninham and Bostr\"om\,\cite{PhysRevA.67.030701,EPJDNinham2014}, though without calculation of lifetimes, following from on Ninham's earlier theoretical calculation of plasmon lifetimes, a substantial work confirmed experimentally by Powell and Swanson\,\cite{NinhamPhysRev.145.209}. Lifetime calculations of muons were addressed more recently in an article by Ninham, Brevik and Bostr\"om in Substantia\,\cite{Ninham_Brevik_Bostrom_2022}.

As Barry Ninham, Adrian Parsegian, and George Weiss\,\cite{NinhamParsegianWeiss1970} showed in 1970,  one can derive  Lifshitz interactions  across an intervening medium from semi-classical theory. It was supposed, along with the Lamb shift,  to represent the pinnacle and confirmation of quantum electrodynamics -- within the assumption about bulk medium up to a molecular distance from an interface. But how  could it be derived from semi-classical theory? What was going on? What was going on was a sleight of hand whereby at a certain point in the formalism of Lifshitz theory, more correctly identifed as the DLP theory of Dzyaloshinskii, Lifshitz and Pitaevskii\,\cite{Dzya}, one had to solve a so-called Dyson nonlinear integral equation for the dielectric susceptibility involving a coupling constant integral. DLP linearised this, rendering the whole thing analytical---and collapsing it to semi-classical theory, not the full QED embodied in the integral equation. So then it becomes interesting. Because IF the semi-classical theory is shown to be equivalent to the theory of weak interactions in particle physics, then we must conclude, or at least suspect, that particle physics theory is equally deficient. The particle physics theory is similarly built on a linear foundation, and is potentially equally flawed. If one dismisses the equivalence, then one has to ask where on earth this energy from the foundations of quantum electrodynamics went to. The problem is too profound to be dismissed.

Philosophically, we know---after Dyson's paper on Feynman's derivation of Maxwell's equations---that  only causality and the finite velocity of light are involved, and the imposition of Planck quantisation is all that is required. We can seek further illumination by including magnetic susceptibility and bringing in developments in non-linear optics. Clearly there is much work still to do. Ironically, the linearisation of the coupling constant integration led to an early version of DLVO theory\,\cite{DrewNote}. 
This was published by Sam Levine \cite{Levine1939_I,Levine1939_II}.  The correct theory was done by Derjaguin with Landau in Russia in 1941 \cite{DeryaguinLandau1941} and by Overbeek \cite{VerweyOverbeek1948} working at Philips industries in occupied Holland during the war, protected by Verwey. Prior to his landmark 1941 paper with Landau, Derjaguin had already commented that Levine's error was a mishandling of the thermodynamic independence of ion charge from ion density profiles \cite{Derjaguin1940}. Levine stubbornly resisted correction, asserting the error was Derjaguin's (and Langmuir's) \cite{LevineDube1939}, although he started to acknowledge the need for a broader thermodynamic perspective during a war-time Faraday meeting of the Royal Society of Chemistry \cite{FaradayTrans1940}. Levine's stubbornness was not well received by Derjaguin and Landau, who caustically referred to Levine's stance as a missed opportunity made ``who knows why?'' \cite{DeryaguinLandau1941rus} (the English translation \cite{DerjaguinLandau1941_1993eng} does not fully convey the frustration emoted in the original text). It took Overbeek's private communications to convince Levine that the term he had missed was the $\ln c$ entropy of the ion density profile \cite{Levine1946} (or osmotic energy \cite{Levine1951}). It is rather ironic that Landau totally dismissed Levine's work because a linearisation assumption, which is to say, a mistake,  was used by  Dzyaloshinskii, Lifshitz and Pitaevskii in their general theory of van der Waals forces \cite{DzyaloshinskiiLifshitzPitaevskii1961}. Levine learnt his lesson well, correcting Babchin's linearisation error \cite{Babchin1974}  three decades later on the theory of electrostriction \cite{BellLevine1976}, albeit without acknowledging Derjaguin's equivalent criticism on the same question \cite{Derjaguin1975_1992}. It is the same problem in classical statistical  mechanics solving the Ornstein-Zernicke equations connecting the indirect and direct correlation functions, with a linear coupling constant coupling polarisation and density fluctuations in the OZ Hamiltonian \cite{KornyshevLeikinSutmann1997}. In this case a sufficiently large coupling leads to divergent oscillatory nonlocal electrostatics, requiring additional coupling terms to regularise energies \cite{BasilevskyParsons-JCP98-osc}.

 In accordance with Substantia's purview to promote the history of science, we present the  work ``Observations on the possible electromagnetic nature of nucleon interactions and pions'' by Barry Ninham and Colin Pask, as a subject of historical interest. The original text commences here.

\section*{Observations on the possible electromagnetic nature of nucleon interactions and pions\\[2\baselineskip]
 Barry Ninham and Colin Pask, 1969}

\section{Introduction}
\label{intro}

At the present time there is much interest, both theoretical and experimental, in the theory of nucleons and related particle. It is not our intention to survey this field in detail. For example of recent papers see Ratner et al.\,\cite{Ratner} 
and Amaldi et al.\,\cite{Amaldi}. Survey-discussion articles by Lee\,\cite{Lee} and Drell\,\cite{Drell} could be consulted, while the article by Kendall and Panofsky\,\cite{Kendall} is a non-technical survey. This is just a further step in the process which began with the explanation of the structure and properties of macroscopic systems in terms of molecules and atoms, continued with the theory of atoms as collections of electrons and muclei, and then scaled down even further to the description of nuclear phenomena as manifestations of the existence and behaviour of nucleons. Recent high-energy experiments, particularly those using the SLAC electron beam, are now forcing theorists to regard nucleons as extended objects, and to reevaluate the old meson cloud model\,\cite{Galloway} or to look for new models. The constituents of the required new model are as yet unknown, but Feynman's generic term ``parton'' is widely used.

To the physicist (as against the mathematician or mathematical physicist) a theory of the structure and interaction of particles seems to be long overdue. Without an underlying model the work of recent years has become more and more mathematical, with the express aim of using as few of the most general physical principles as possible. Such
principles might be relativistic invariance, causality and known conservation laws, giving rise to dispersion relations and unitary symmetry for example. The theory of quarks offered some hope, for as Weisskopf\,\cite{Weisskopf} has written: `` \ldots it serves as a simply describable realization of $\mathrm{SU}(3)$ symmetry. The latter is what remains of the quark model, if one removes the quarks - the grin of the Cheshire cat''. Quarks may be partons, but their existence is as yet uncertain.

In this paper we do not consider the detailed nature of nucleon structure, but speculate on how that structure may enable \underline{electromagnetic} theory to be built into strong and weak interaction theories as a major and unifying concept. (We note that current algebra strongly suggests such a close link, based however on plausible mathematics rather than on an underlying physical model. See Weisskopf\,\cite{Weisskopf}, for example.)

In one sense we are suggesting that before postulating new mechanisms such as parton-parton forces, the known forces should be tried first. We do not exclude new types of forces which may also be of importance in explaining certain problems. A similar (but \emph{not} analogous) situation occurs in alpha particle scattering: at low energies the $\alpha$-$\alpha$ force will be electromagnetic, while the $\alpha$ particles themselves are held together by other forces. These other forces become of importance in the high energy - small distance region. In atom-atom interactions, on the other hand, the interaction and structure forces both have the same origin.

As a further general point we remark that we continue to use the usual quantum mechanics even in the sub-nucleon limit. The validity of this assumption is of course unknown, but quantum electrodynamics does appear to work at very small distances.

In section \ref{section2} we discuss the possibility of electromagnetic fluctuation forces being responsible for strong interactions. We assume further that at the very high photon frequencies involved there is an equilibrium between the photons and electron positron pairs as discussed by Landau and Lifshitz\,\cite{LandauLifshitzStatPhys1}. This introduces a plasma into our theory. Collective excitations may be set up in the plasma and we identify these plasmons with the pions of conventional theory. Such a theory enables the $\pi^{0}$ mass and lifetime to be calculated without introducing conventional strong or weak interaction theories. The details of this theory of the pion are presented in section  \ref{section3}. Concluding remarks are made in section  \ref{section4}.
\\

\section{Electromagnetic theory and nucleon interactions}
\label{section2}

The theory of electromagnetic fluctuation forces acting between macroscopic bodies is by now well established and understood\,\cite{landau1960electrodynamics}, and is now being used at the microscopic level by McLachlan\,\cite{McLachlan}, Ninham and Parsegian\,\cite{NinPars1970,NinhamParsegian1970JCP} and others. Fluctuation forces are of short-range and may be large, and so may be suitable candidates for explaining nuclear forces. In this paper we assume that the nucleons have structure which produces electromagnetic fluctuation forces, and, as a first step, we give some rough energy and distance arguments.

Consider two semi-infinite planes of nuclear material a distance d apart. If we consider perfectly reflecting planes, the electromagnetic energy of interaction per unit area due to the wave system set up in the gap as obtained by Casimir\,\cite{Casi} (see also Power\,\cite{Power}) is
\begin{equation}
    \mathrm{E}_{\text{A}}= - \frac{\pi^{2}}{720} \frac{\hbar c}{d^{3}}
    \label{eqn:15}
\end{equation}

Now consider two protons with their centres d apart. As a first crude approximation, we use the above result with surface area $\pi r_{0}^{2}, r_{0}=$ proton radius $\sim 0.9 \mathrm{f}$. Then the attractive force is $\sim-[\pi^{3} \hbar c r_{0}^{2}] / [240 \mathrm{~d}^{4}$]. A typical nuclear distance might be obtained by balancing this force with the repulsive Coulomb force $e^{2} /\left(2 r_{0}+d\right)^{2}$. This gives $d \sim 5 f$ with a corresponding energy $~ 1 / 5 \mathrm{MeV}$. Obviously one cannot expect too much from
such a crude argument, but we feel that the important point is that the distance is of order fermis rather than \AA{ngstroms} or cm's., and the energy is of order MeV rather than ev or ergs. Somewhat more striking is the observation that the attractive electromagnetic energy between two nucleons at a distance of 1f as computed by the above procedure is $\approx 10 \mathrm{MeV}$! 

To proceed further we would need to assume or deduce details of the nucleon structure, i.e. the electromagnetic susceptibility.

\section{Electromagnetic theory and pions}
\label{section3}

In the standard scalar theory, nuclear interactions proceed via a field described by the Klein-Gordon equation, which in Fourier space may be cast into the form
\begin{equation}
    \nabla^{2} \phi_{\omega}-\frac{\omega^{2}}{c^{2}}\left(1-\frac{\mu^{2} c^{2}}{\omega^{2}}\right) \phi_{\omega}=0 
    \label{eqn:KleinGordon}
\end{equation}
where $\mu= \mathrm{m}_{\pi} c / \hbar, \mathrm{m}_{\pi}=$ pion mass. There is a suggestive similarity between this equation and that for the Maxwell scalar potential. In the high frequency regime we therefore identify $1-\frac{\mu^{2} c^{2}}{\omega^{2}}$ with the electromagnetic susceptibility $\varepsilon(\omega)$ which here has the form appropriate to a plasma. Thus we obtain
\begin{equation}
    \mu^{2} c^{2}=\omega_{\rho}^{2}=\frac{4 \pi {N} \mathrm{e}^{2}}{ \mathrm{m}_{e}}
    \label{eqn:17}
\end{equation}
where $\mathrm{m}_{\mathrm{e}}$ = electron mass, and ${N}$ the electron number density. (The standard plasma theory considers a set of negative charges with a smeared-out positive background to preserve electrical neutrality. Here the positive component of our system is the positron gas and we should actually use a theory of an electron-positron plasma. However no such theory is available in detail, and we believe the conventional theory can be used in any first rough studies).

Eq. (\ref{eqn:17}) contains $\mathrm{m}_{\pi}$ and ${N}$ which are still unknown. We now give arguments which determine these and other important quantities.

\subsection*{Mass of the pion}
Again we consider the nuclear material slabs interacting across a gap of width $d$. The electromagnetic energy in the gap has a density ${E}$ which follows from Eq. (\ref{eqn:15}):
\begin{equation}
    {E}=-\frac{\pi^{2}}{720} \frac{\hbar \mathrm{c}}{\mathrm{d}^{4}}
\end{equation}
The equivalent temperature of this (virtual) energy can be found by saying that the energy involved is the same as a unit volume black box, 

\begin{equation}
    E=\frac{\pi^{2}(\mathrm{kT})^{4} }{15 ( { \hbar c })^{3}}.
\end{equation}
Thus we obtain,
\begin{equation}
    \mathrm{kT}=\frac{\hbar c}{2\times 3^{1 / 4} \mathrm{~d}}.
    \label{eqn:19}
\end{equation}
Actually, we will have a system of photons plus electron-positron pairs. If we assume $d \sim$ 1f the equivalent temperature is so high, $\mathrm{kT} \sim 140 \mathrm{~m}_{\mathrm{e}} \mathrm{c}^{2}$ and the black box will contain almost the maximum number of $\mathrm{e}^{+} \mathrm{e}^{-}$ pairs. We then obtain\,\cite{LandauLifshitzStatPhys1}

\begin{equation}
    N = N^{-} \approx 0.183\times (\frac{\mathrm{kT}}{\hbar c})^{3}
    \label{eqn:20}
\end{equation}

We now take $d \sim$ 1\,fm, calculate $N$ using Eqs. (\ref{eqn:19}) and (\ref{eqn:20}) and hence obtain the pion mass using Eq.(\ref{eqn:17}). This gives
\begin{equation}
    \mathrm{m}_{\pi} \approx 220 \mathrm{~m}_{\mathrm{e}}
    \label{eqn:21}
\end{equation}
which compares favourably with the experimental result $m_{\pi} \simeq 270 \mathrm{~m}_{e}$.

\subsection*{The neutral pion lifetime}

The pion plays a key role in the usual theory of nuclear forces, e.g. Wick\,\cite{Wick} used its mass to predict the force range. We have already shown how range, energy considerations and the pion mass are connected in our theory. We can now proceed to a further property of the pion by using our assumption that the collective excitations in the plasma produced by the electromagnetic fluctuations are to be identified with pions. This is simplest for $\pi^0$. Now although plasma theory is at present far from complete, we believe that the plasma lifetime calculations given by Ninham\,\cite{Ninham1966} should give a correct order of magnitude for the present case. The lifetime mentioned is for the decay plasmon $\rightarrow$ two e$^+$e$^-$ pairs. In the sea of e$^+$e$^-$ pairs, these can decay into photons. Using Eq.\,(10) given by Ninham\,\cite{Ninham1966}, with q associated with m$_\pi$c$^2$, we obtain

\begin{equation}
    \pi^{0} \,\,\mathrm{lifetime}\,\, \leq 1.5\times10^{-17}\,\mathrm{s}.
\end{equation}
The current experimental value is $(0.84\pm 0.1)$ $\times10^{-16}$\,s\,\cite{Rittenberg}.

Clearly this theory predicts various decay modes for $\pi^0$, all in terms of e$^+$e$^-$ pairs and photons, again in agreement with experiment. The coupling of the neutral pion to the electromagnetic field raises no problems in this theory. In the older conventional theory an intermediate proton-anti-proton state is introduced, $\pi^0\rightarrow p \overline{p}\rightarrow 2\gamma$\,\cite{Finkelstein}, and thus the coupling is via strong interactions. The $\pi^0$ lifetime may then be related to proton Compton-scattering as in Jacobs and Mathews\,\cite{Jacob}. More recent theories seem unable to account for $\pi^0$ decay and it remains one of the three ``unresolved'' problems of current algebra, according to Beg\,\cite{Beg}. We note that our result does not involve relations with other processes and does not contain any strong interaction coupling constant.

\subsection*{Plasmon critical wavenumber}
There is a critical wave number $\mathrm{k}_{c}$ for plasmon existence\,\cite{Raimes}. Clearly, this is related to the interaction energy of nucleons and pion energies and masses in this theory. However, the theory for $k_{c}$ is still uncertain, and so, at this stage, we do not attempt to use this effect in any detail.\\

\section{Conclusions}
\label{section4}
We do not claim any great accuracy for the above calculations, but rather, we take the point of view that our rough ideas could lead to an interesting quantitative theory if correctly interpreted. For example, the plasma theory for the pairs and densities we postulate would need to be investigated. 

As regards the description of particles other than $\pi^{0}$, several possibilities may be considered. Associating charge with a plasmon could give $\pi^{\pm}$. Decay of charged pions could reveal the role of the muon in particle theory, and of course, this would then bring weak interactions into the theory. One could also consider a generalized vector Klein-Gordon equation with a ``magnetic susceptibility'' term in analogy with the electromagnetic vector potential equation. 

Finally, we point out that we consider basically nuclear interactions and not ``free particles in a vacuum'' since any ``observed vacuum'' is always bounded by interacting matter. (However, we note that all electron theories from Dirac's negative energy states sea onwards have contained vacuum fluctuations.  Hence, a sea of  (virtual) $\mathrm{e}^{+}\mathrm{e}^{-}$ pairs (as in our theory) is associated even with the vacuum). Thus our ideas are in line with recent attempts to use only real particles and to avoid the difficulties associated with a description based on free, bare particles with interactions then introduced via perturbation theory.


\begin{acknowledgments}
MB's contributions in the current work is part of the project No. 2022/47/P/ST3/01236 co-funded by the National Science Centre and the European Union's Horizon 2020 research and innovation programme under the Marie Sk{\l}odowska-Curie grant agreement No. 945339.  Institutional and infrastructural support for the ENSEMBLE3 Centre of Excellence was provided through the ENSEMBLE3 project (MAB/2020/14) delivered within the Foundation for Polish Science International Research Agenda Programme and co-financed by the European Regional Development Fund and the Horizon 2020 Teaming for Excellence initiative (Grant Agreement No. 857543), as well as the Ministry of Education and Science initiative “Support for Centres of Excellence in Poland under Horizon 2020” (MEiN/2023/DIR/3797). MB gratefully acknowledge the kind hospitality by Assoc. Prof. Drew Parsons and the University of Cagliari (Italy) where this work was partly prepared. We finally thank Mr Subhojit Pal for his help typing up the original scanned version of the manuscript.
\end{acknowledgments}

\begin{figure}[ht]
  \centering
\includegraphics[width=\columnwidth]{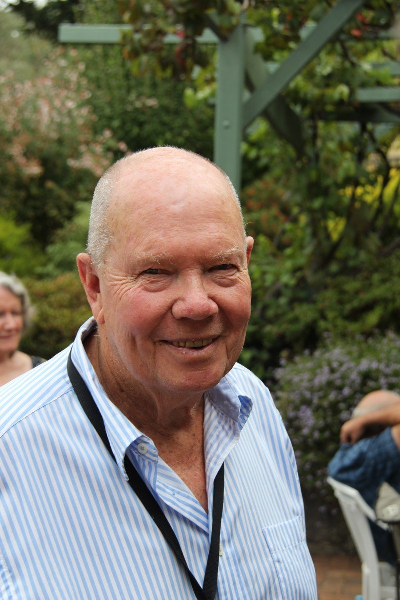}
  \caption{\label{scheme} Graphical Abstract: Barry Ninham, founder of the Department of Applied Mathematical physics at the Australian National University in 1970. }
\end{figure}


\begin{thebibliography}{69}
\bibitem [{Dre()}]{DrewNote}%
  \BibitemOpen
  \href@noop {} {}\bibinfo {note} {{A note from Drew F. Parsons: I am not certain that linearisation of the coupling constant was the problem with Levine's pre-DLVO. The issue was a mishandling of the independence of thermodynamic variables, not linearisation of the coupling constant, at least not explicitly.}}\BibitemShut {Stop}%
\end{thebibliography}
%
\end{document}